\documentclass[twocolumn,aps,prb,epsf,showpacs,superscriptaddress]{revtex4}
\usepackage{amsmath}
\usepackage{amsfonts}
\usepackage{epsfig}
\usepackage{epsf}
\usepackage{array}
\usepackage{color}

\begin{document}

\title{Spontaneous fourfold-symmetry breaking driven by electron-lattice coupling and strong correlations
in high-$T_c$ cuprates}

\author{Satoshi Okamoto}
\altaffiliation{okapon@ornl.gov}
\affiliation{Materials Science and Technology Division, Oak Ridge National Laboratory, Oak Ridge, Tennessee 37831, USA}
\author{Nobuo Furukawa}
\affiliation{Department of Physics and Mathematics, Aoyama Gakuin University, Sagamihara 229-8558, Japan}

\begin{abstract}
Using dynamical-mean-field theory for clusters, 
we study the two-dimensional Hubbard model in which 
electrons are coupled with the orthorhombic lattice distortions through the modulation in the hopping matrix. 
Instability towards spontaneous symmetry breaking from a tetragonal symmetric phase to 
an orthorhombic distorted phase is examined as a function of doping and interaction strength. 
A very strong instability is found in the underdoped pseudogap regime 
when the interaction strength is large enough to yield the Mott insulating phase at half filling. 
The symmetry breaking accompanies the recovery of quasiparticle weights along one of the two antinodal directions, 
leading to the characteristic Fermi arc reconnection. 
We discuss the implications of our results to the fourfold symmetry breaking reported in systems 
where the underlying crystal does not have any structural anisotropy. 
\end{abstract}

\maketitle




Electronic nematicity has become one of the central subjects of correlated-electron systems.%
\cite{Yamase00,Halboth00,Kivelson03,Fradkin09,Borzi07,Chu10} 
For high-$T_c$ cuprates, very large anisotropies in low energy excitations 
have been experimentally reported,\cite{Ando02,Lee02,Hinkov08,Daou10,Lawler10,Haug10,Kohsaka12} 
and their connection with the ``pseudogap phase'' has been discussed.

In a system such as YBa$_{2}$Cu$_{3}$O$_{6+x}$ (YBCO), 
there exists an intrinsic structural anisotropy resulting in a tiny but finite band-structure anisotropy. 
This band anisotropy has been shown to induce huge effects in low-energy excitations\cite{Yamase06}
when the system is close to a correlation-induced Pomeranchuk instability.\cite{Halboth00}
Although cluster dynamical mean-field studies do not find a spontaneous symmetry breaking 
in a two-dimensional single-band Hubbard model,\cite{Gull09,Okamoto10} 
a tiny band anisotropy was shown to dramatically amplify the anisotropy 
in the dc transport and electronic excitation spectrum 
in the underdoped pseudogap regime.\cite{Okamoto10,Su11}
On the other hand, in a system such as Bi$_2$Sr$_2$Ca$_{n-1}$Cu$_n$O$_{8+y}$ (BSCCO) and 
Ca$_{2-x}$Na$_x$CuO$_2$Cl$_2$ (CNCOC), 
there is no intrinsic structural anisotropy 
but the symmetry breaking from tetragonal ($C_4$) to orthorhombic ($C_2$), called intra-unit-cell (IUC) nematicity,
has been observed.\cite{Lawler10,Kohsaka12} 
Within a mean-field treatment, IUC order in the Emery model for the CuO$_2$ plane has been analyzed.\cite{Fischer11} 
However, very large interactions are required to realize the IUC symmetry breaking. 
This may indicate the importance of additional degrees of freedom.

Here, we consider a correlated model for cuprates including the coupling between electrons and lattice distortions (EL) 
as a possible ingredient for experimentally reported spontaneous $C_4$ symmetry breaking. 
We observed a moderate tendency towards symmetry breaking when the chemical potential is located 
near the van Hove singularity. 
In addition, we found a very strong instability in the underdoped pseudogap regime when 
the interaction strength is large enough to yield a Mott insulating state at half filling. 
The stabilization of the distorted phase comes from the gain in the kinetic energy. 
In the overdoped regime, the Fermi surface is deformed to split the van Hove singularity at $(\pi,0)$ and $(0,\pi)$ 
and shift it from the Fermi level, 
while in the underdoped regime in the presence of strong Coulomb interaction 
the pseudogap becomes anisotropic. 
Our results may provide a coherent picture connecting the ``electron nematicity'' and pseudogap behavior in 
high-$T_c$ cuprates highlighting the difference between 
YBCO, whose structure is intrinsically anisotropic, 
and BSCCO, which is structurally isotropic but the $C_4$ symmetry 
is found to be locally broken.

\begin{figure}[tbp]
\includegraphics[width=0.7\columnwidth]{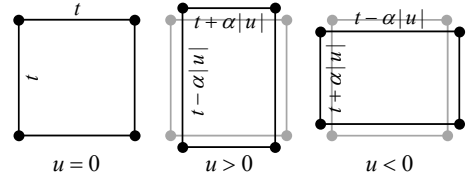}
\caption{Model electron-lattice coupling. 
Orthorhombic distortion $u$ induces the hopping anisotropy with the coupling constant $\alpha$.}
\label{fig:distortion}
\end{figure}

We consider the following electron-lattice coupled model: $H=H_{ele}+H_{latt}$. 
$H_{ele}$ is the two-dimensional (2D) Hubbard model as a generic model for high-$T_c$ cuprates: 
\begin{equation}
H_{ele}=-\sum_{ij\sigma }t_{ij}d_{i\sigma }^{\dag }d_{j\sigma} + U\sum_{i} d_{i
\uparrow}^\dag d_{i \uparrow} d_{i \downarrow}^\dag d_{i \downarrow}.
\label{eq:Hhub}
\end{equation}%
Here, $d_{i\sigma }$ is the annihilation operator for an electron with spin $\sigma $ at site $i$
and $U$ is the local Coulomb interaction. 
The band structure part is described by $t_{ij}$; 
$t_{i \neq j}$corresponds to the transfer integral and  
and $t_{i=j} = \mu $ is the chemical potential. 
The lattice part is given by $H_{latt} = \frac{NK}{2} u^2$, 
where $u$ is the orthorhombic distortion, $K$ is the elastic constant, and $N$ is the total number of sites. 
We consider the coupling between the orthorhombic distortion and electrons 
as the modulation in the nearest-neighbor (NN) transfer integral $t$ along the $x$ and $y$ directions as 
$t_{x,y}=t \pm \alpha u$ as illustrated in Fig.~\ref{fig:distortion}, 
with the next-nearest-neighbor transfer integral $t^{\prime }$ unaffected by the distortion. 
This electron-lattice coupling may also be regarded as a simplified one 
realized in, for example, a low-temperature tetragonal (LTT) phase. 
Here, the crystal structure is ``tetragonal'' with the equal lattice constants along the $a$ and $b$ axes 
but the electronic band structure is ``orthorhombic'' 
due to the coherent rotation of CuO$_6$ octahedra.\cite{Axe89,Nachumi98,Yamase00}

We analyze our model using the cellular dynamical-mean-field theory (CDMFT) 
(Refs.~\onlinecite{Kotliar01} and \onlinecite{Kotliar06}) at zero temperature. 
This method captures the full dynamics [i.e., the frequency dependence of the spectral function (SF)] 
and the short-ranged spatial correlations beyond the single-site dynamical-mean-field theory
and has been applied for a variety of problems in low-dimensional systems.\cite{Capone04,Kyung06,Kancharla08}
The CDMFT maps the bulk lattice problem onto an effective Anderson model 
describing a cluster embedded in a bath of noninteracting electrons. 
The short-ranged dynamical correlations are treated exactly within the cluster. 
In this study, we employ a $2\times 2$ plaquette ($N_c=4$) coupled to eight bath orbitals and solve 
it using the Lanczos exact diagonalization technique\cite{Caffarel94} 
which requires a low-energy cutoff $\omega_c$ corresponding to 
the discrete imaginary frequency as $\omega_n = (2n+1) \omega_c$. 
In this work, we take $\omega_c = 2 \times 10^{-2} \pi t$. 
$\omega_c$ should not be interpreted as real temperature (times $\pi$) 
because only the ground state of the impurity model is taken to compute the Green's function. 
The numerical details are described in Ref.~\onlinecite{Okamoto10}.

In the following discussion, 
we use as a band parameter $t'=-0.3t$ that is appropriate for cuprates. 
For interaction strength, we consider $U=10t$, $4t$ and $0$. 
The largest $U$ is supposed to be relevant for cuprates.

We start from the discussion on the first instability caused by the EL coupling. 
This could in principle be done by computing the $C_2$ susceptibility 
and finding a parameter range where the susceptibility diverges. 
Such an analysis normally requires the inclusion of vertex corrections, 
but the precise form is unknown for the current CDMFT technique. 
Instead, we utilize the Ginzburg-Landau theory with 
the self-energy functional approach \cite{Potthoff03} by which 
the electronic contribution $F_{ele}$ to the free energy $F_{tot}=F_{ele}+F_{latt}$ is written as
\begin{eqnarray}
F_{ele} \!\!&=&\!\!  
- T \sum_{\omega_n} \int_{\tilde{\mathbf{k}}} \ln \det 
\bigl\{1 - \bigl[\hat t(\tilde{\mathbf{k}}) - \hat t^{c} - \hat \Gamma (i \omega_n) \bigr] 
\hat G^{c} (i \omega_n) \bigr\} \nonumber \\
&&+ F_{c}. \label{eq:free}
\end{eqnarray}
Here 
$t^{c}_{ij} = t_{ij}$ is the hopping matrix on the $2 \times 2$ plaquette, and 
$t_{ij}(\widetilde{\mathbf{k}})=N_{c}^{-1}\sum_{\mathbf{K}}
e^{i\left( \mathbf{K}+ \widetilde{\mathbf{k}}\right) \cdot \left( \mathbf{r}_{i}-\mathbf{r}_{j}\right)}
\varepsilon _{\mathbf{K}+\widetilde{\mathbf{k}}}$ 
describes the hopping between the clusters covering the original lattice. 
$\widetilde{\mathbf{k}}$ are wave vectors in the reduced Brillouin zone, 
and $\mathbf{K} = (0,0),(\pi,0),(0,\pi)$, and $(\pi,\pi)$. 
The bare dispersion is given by 
$\varepsilon _{\mathbf{k}}=-2(t_{x}\cos k_{x}+t_{y}\cos k_{y}+2t^{\prime }\cos k_{x}\cos k_{y})$. 
$\hat \Gamma (i \omega_n)$ is the hybridization function with which the cluster Green's function at the ground state 
is written as 
$\hat G^{c} (i \omega_n) = [i \omega_n - \hat t^{c} - \hat \Gamma (i \omega_n) - \hat \Sigma (i \omega_n)]^{-1}$, 
where $\hat \Sigma (i \omega_n)$ is the cluster self-energy. 
Because of the low-energy cutoff $\omega_c$, 
the free energy is approximately calculated using the finite-temperature form Eq.~(\ref{eq:free}) at 
$T = \omega_c/\pi$. 
As a result, the results are somewhat smeared out, underestimating the instability. 
Finally, $F_{c}$ is the free energy of the cluster model. 
The lattice contribution is given by $F_{latt}=\frac{N_c}{2} K u^2$.

In practice, 
we compute $F_{ele}$ as a function of the hopping anisotropy $\delta_t = \alpha u$ as 
$\Delta F_{ele} (\delta_t) = F_{ele}(\delta_t) - F_{ele}(0) \approx - \frac{N_c}{2} \beta \delta_t^2 + {\cal O}(\delta_t^4)$. 
Normally, the ${\cal O}(\delta_t^4)$ contribution is positive. 
As the lattice contribution $F_{latt} \rightarrow \frac{N_c}{2 \alpha^2} K \delta_t^2$ is always positive, 
quadratic fitting to $\Delta F_{ele}(\delta_t)$ gives 
the critical EL coupling towards the spontaneous $C_4$ symmetry breaking as $\alpha^2/K=1/\beta$. 
The critical point deduced in this way signals the second-order transition. 
When the ${\cal O}(\delta_t^4)$ contribution is negative, the transition becomes first order, 
thus, $\alpha^2/K=1/\beta$ should be regarded as the upper limit of the coupling 
above which a distortionless state is no longer an energy minimum. 
Instead, the critical EL coupling for the first-order transition is located at the smallest $\alpha^2/K$ which satisfies 
$\Delta F_{ele}(\delta_t) + \frac{N_c}{2 \alpha^2} K \delta_t^2 = 0$ at $\delta_t \ne 0$. 
Because of the higher-order terms ${\cal O}(\delta_t^4)$ and ${\cal O}(\delta_t^6)$ in $F_{ele}$ and 
the small change in the carrier density with $\delta_t$, 
these fitting procedures are notoriously difficult. 
Nevertheless, an overall trend can be deduced.

\begin{figure}[tbp]
\includegraphics[width=0.9\columnwidth,clip]{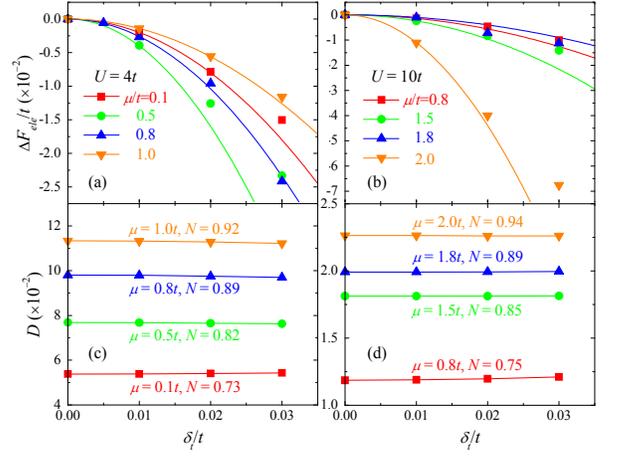}
\caption{(Color online) $\delta_t$ dependence of electronic free energy $\Delta F_{ele}$ (a) and (b) 
and the double occupancy $D$ (c) and (d). 
$U=4t$ for (a) and (c), and $U=10t$ for (b) and (d). 
Solid lines in (a) and (b) are quadratic fits to the numerical data of $\Delta F_{ele}$. 
}
\label{fig:deltat}
\end{figure}

As examples, we plot $\Delta F_{ele} (\delta_t)$ in Figs.~\ref{fig:deltat}(a) and 2(b). 
$\Delta F_{ele}$ decreases with increasing $\delta_t$. 
According to the linear combination of atomic orbitals method,\cite{Harrison80,footnote}
the EL coupling constant $\alpha$ is estimated to be 0.9~eV/{\AA}, 
and the distortion considered here is rather small; 
$\delta_t =0.03$ corresponds to $u \sim 0.033$~{\AA}. 
In most cases, $\Delta F_{ele}$ deviates from the quadratic curve upwards. 
While for $U=4t$ with $\mu=0.8t$ and $U=10t$ with $\mu=1.8t$, 
$\Delta F_{ele}$ shows a small downward deviation, indicating first order transitions. 
For $U=10t$ with $\mu=2.0t$, corresponding to the underdoped regime. 
$\Delta F_{ele}$ shows a strong dependence on $\delta_t$, indicating strong instability.

As shown in Fig.~\ref{fig:deltat}, the $\Delta F_{ele}$-$\delta_t$ curve is rather sensitive to the doping concentration. 
To see its origin, we plot, in Figs.~\ref{fig:deltat}(c) and 2(d), the double occupancy 
$D=\langle d_{i \uparrow}^\dag d_{i \uparrow} d_{i \downarrow}^\dag d_{i \downarrow} \rangle$
as a function of $\delta_t$. 
One notices that the double occupancy remains unchanged, 
i.e., the potential energy $UD$ remains unchanged. 
Thus, the gain in the ``kinetic energy'' dominates the $\Delta F_{ele}$-$\delta_t$ 
irrespective of doping dependence.

\begin{figure}[tbp]
\includegraphics[width=0.7\columnwidth,clip]{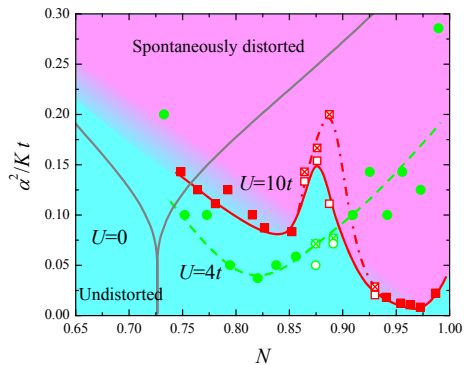}
\caption{(Color online) Phase diagram for the 2D Hubbard model with the EL coupling 
as a function of electron density $N$ and the coupling constant $\alpha^2/K$. 
Parameters are $U=10t$ (squares) and $U=4t$ (circles) with $t^{\prime }=-0.3t$. 
The critical points for the second- (first-) order transition are indicated by filled (open) symbols. 
The critical points where distortionless states lose metastability are indicated by crossed symbols. 
Thick lines are guides to the eye. 
For comparison, the phase boundary for $U=0$  
is shown as a light solid line.
}
\label{fig:diagram}
\end{figure}

Figure \ref{fig:diagram} shows the resulting phase diagram.\cite{Note}
As described below, instabilities appear at two doping regimes due to different mechanisms, 
say type A and type B. 
%
Type A is a weak-coupling mechanism and appears 
for both $U=4t$ (boundary is indicated by circles) and $U=10t$ (squares) 
at $N\sim0.8$, near the van Hove filling. 
Type B, on the other hand, is a strong-coupling mechanism and only appears when $U$ is large 
in the underdoped regime ($N>0.9$ for $U=10t$).

\begin{figure}[tbp]
\includegraphics[width=0.9\columnwidth]{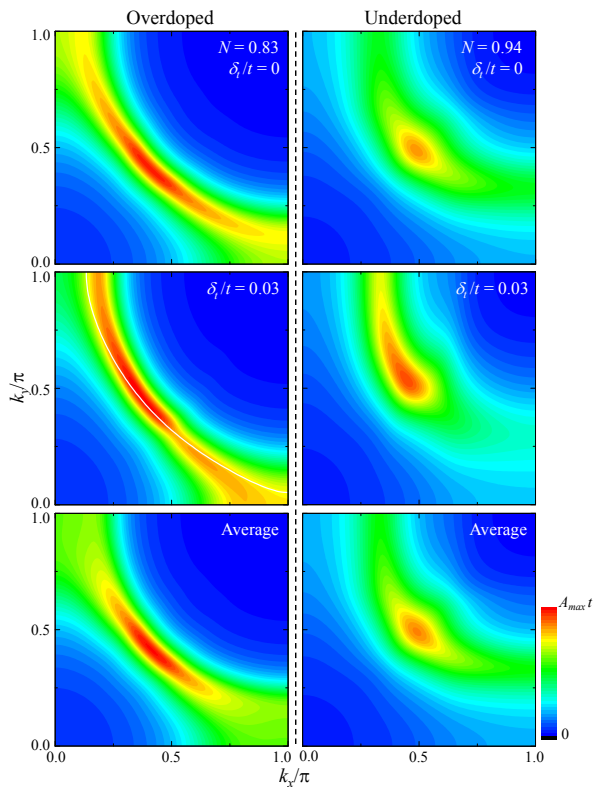}
\caption{(Color online) Evolution of the spectral function at the Fermi level 
in the first quadrant of the Brillouin zone for $U=10t$. 
Left panels: results for $N=0.83$ (overdoped regime), and right panels: results for $N=0.94$ (underdoped regime). 
Top: Symmetric SF; middle: asymmetric SF with $\delta_t/t=0.03$; bottom: averaged SF with $\delta_t/t = \pm 0.03$. 
The Green's-function periodization scheme is used 
with the small imaginary part $i \eta$ $(\eta=0.1t)$ added to the real frequency. 
The maximum of the spectral weight is $A_{max}t = 0.6$ (left) and 0.35 (right). 
A white line in the middle-left panel shows the FS for $\delta_t/t=0.03$, $N=0.83$ and $U=0$. 
}
\label{fig:FS}
\end{figure}

Type-A weak coupling instability also appears for $U=0$. 
In this case, the critical coupling is given by expanding the free energy up to ${\cal O}(\delta_t^2)$ 
and equating its coefficient to zero, and the resulting expression is 
$\alpha^2/K = - \pi^2 / 2 \int dk^2 \bigl(\cos k_x-\cos k_y\bigr)^2 f' \bigl(\varepsilon_{\mathbf{k}} - \mu\bigr)$, 
where $f'$ is the derivative of the Fermi-Dirac distribution function. 
Due to the logarithmic divergence in the DOS, 
the critical coupling is minimized at the van Hove filling $N \sim 0.726$ 
as shown as a light solid 
line. 
By finite $U$, the instability is shifted to larger $N \sim 0.8$. 
From the analysis of SFs, the shift in the critical $N$ is caused by the enhanced band anisotropy 
due to correlations. 
The corresponding SFs [contour plot of the SF, 
$A(\mathbf{k},\omega =0) = - \frac{1}{\pi} {\rm Im} G(\mathbf{k},\omega =0)$] 
are presented in Fig.~\ref{fig:FS} (top left, isotropic band) and (middle left, anisotropic band), 
where the Green's function periodization scheme is adopted.\cite{Senechal00} 
The enhanced band anisotropy is evident from the comparison with 
the FS for $U=0$ (a white line). 
By finite $U$, the FS opens up to become quasi-one-dimensional [see Fig.~\ref{fig:FS} (middle left panel)]. 
This is favorable for gaining the kinetic energy 
by splitting the van Hove singularity and shifting it from the Fermi level.
This could also explain why the first-order transition appears at $N > 0.8$, 
where correlation effects are stronger.\cite{Khavkine04}

For the type-B instability, the band anisotropy remains almost unchanged but 
the anisotropy in the scattering rate is enhanced significantly. 
These points can be clearly seen in the corresponding SFs presented in Fig.~\ref{fig:FS} 
(top right, isotropic band, and middle right, anisotropic band). 
Here more importantly, the coherence is recovered near $(0,\pi)$ by the band anisotropy 
because the FS goes away from the so-called ``hot spot,'' while the FS 
near $(\pi,0)$ approaches the hot spot. 
This results in the reconnection of the ``Fermi arc'' between the first and the second Brillouin zones 
neighboring along the $y$ direction; 
see Fig.~\ref{fig:FS} (middle right) and Fig.~1 (d) in Ref.~\onlinecite{Okamoto10}. 
The relation between the quasiparticle coherency and the kinetic energy 
can be directly seen from the expression for the kinetic energy 
$E_{kine} = T \sum_{\mathbf{k}, n} \varepsilon_{\mathbf{k}} G(\mathbf{k}, i \omega_n)$. 
As the quasiparticle coherency is lost in the ``symmetric'' underdoped pseudogap regime, 
the gain in the quasiparticle weight leads 
the dramatic gain in the kinetic energy as seen in Fig. \ref{fig:deltat} (b), 
resulting in the strong instability in this regime. 

Given the above discussion, the type-B instability is expected to be more relevant for underdoped high-$T_c$ cuprates 
compared with the type-A instability and other weak-coupling instabilities.
The type A instability is suppressed by correlations because the van Hove singularity is smeared out by 
the imaginary part of the self-energy. 
In fact, for $U=10t$, this instability is almost diminished. 
Not only by the imaginary part of the self-energy, 
the type-A instability is also suppressed by finite temperature. 
Because of the linear $T$ dependence of the electronic free energy, 
the instability is expected to go away rather quickly at elevated temperatures. 
On the other hand the type B instability requires large $U$, 
resulting in the pseudogap or the Fermi arc which sets its energy scale. 
Therefore, the type B instability is expected to survive at relatively high temperatures 
as long as the pseudogap or Fermi arc remains. 
%
%
For $U=10t$, there appears a dip in the instability at $N \sim 0.88$ because both instabilities are weak (see Fig.~\ref{fig:diagram}) 
at this doping. 
This does not contradict the seemingly stronger instability near 1/8 doping reported experimentally \cite{Laliberte11} 
as it has different origins. 
Additional weak coupling instability incompatible with our model could take place at carrier densities smaller than $N \sim 0.9$ 
as discussed, for example, in Refs.~\onlinecite{Markiewicz10} and \onlinecite{Markiewicz12}.


When the system is in the vicinity of the structural transition, 
dynamical or statistical fluctuation effects should play important roles. 
If the transition is of the first order, such a critical regime would be characterized by 
a superposition between two distortion modes which minimize the free energy. 
Furthermore, even in a distorted phase, a sample could form domains. 
As a result, low-spatial-resolution angle-resolved photoemission spectroscopy (ARPES) measurements 
would detect the SF that is an average of SFs over a finite lattice spacing. 
We simulate the latter two cases by taking the average of the SFs with different anisotropy parameters, 
$\delta_t/t=\pm 0.03$. 
Figure~\ref{fig:FS} (bottom left and bottom right) show the averaged SF for 
$N=0.83$ (the overdoped regime) and $N=0.94$ (the underdoped regime), respectively. 
For the overdoped regime, the spectral function is broadened at $(\pi,0)$ and $(0,\pi)$ relative to 
the results without distortion. 
Thus, the ``lattice fluctuation'' and structural domain formation act as if enhancing the pseudogap behavior. 
A similar effect has been reported for the thermal nematic fluctuation.\cite{Yamase12}
On the other hand, for the underdoped regime, 
the SF in the undistorted phase and the averaged one in the distorted phase are nearly identical. 
This is because the shape of the FS is insensitive to the band anisotropy in this doping regime. 

In contrast to YBCO, 
BSCCO, La$_{2-x}$Sr$_x$CuO$_4$ and CNCOC do not have a source for band anisotropy, 
yet (local) $C_4$ symmetry breaking has been reported. 
For BSCCO, recent scanning micro-x-ray-diffraction \cite{Poccia11} and 
scanning tunneling microscopy \cite{Zeljkovic12} measurements revealed 
that the system is structurally inhomogeneous involving LTT-like distorted regions. 
These results are consistent with our picture if the EL coupling is in the range of the spontaneous distortion. 
In fact, from the measured elastic constants 
$c \sim 1.7 \times 10^{12}$~dyn/cm$^2$ for La$_2$CuO$_4$ (Ref.~\onlinecite{Migliori90}) 
and $c \sim 1.3 \times 10^{12}$~dyn/cm$^2$ for BSCCO (Ref.~\onlinecite{Wu90}), 
our elastic constant is estimated as $K \sim 6$~eV/{\AA}$^2$ for both systems. 
The resulting EL coupling constant\cite{footnote} $\alpha^2/Kt \sim 0.2$ locates these systems inside 
the spontaneous distortion regime. 
Thus, it is desirable to experimentally clarify the relation between the electronic\cite{Lawler10,Kohsaka12} 
and the structural $C_4$ symmetry breaking.

Our prediction can be tested by 
high-spatial-resolution ARPES measurements as in the x-ray-diffraction measurements in 
Ref.~\onlinecite{Poccia11}. 
For underdoped cuprates below the pseudogap temperature, 
we expect that 
anisotropic FSs as in the lightly doped YBCO (Ref.~\onlinecite{Fournier10})
or reconnected Fermi arcs are spatially distributed. 
The pseudogap behavior should preempt or accompany 
the local lattice distortion 
with small effects on the ``bulk'' SF at the structural transition. 
In contrast, 
the opening of a pseudogap and the local lattice distortion are expected to take place simultaneously in the overdoped regime. 
From these experimental tests, 
a variety of anomalies in relation to the electronic nematicity and pseudogap behavior in high-$T_c$ cuprates 
could be coherently understood in terms of 
the EL coupling and the absence/presence of the intrinsic band anisotropy. 
Further, Raman scattering would be a useful tool to distinguish different roles 
played by electronic systems and lattice (phononic) systems\cite{Lin10,Yamase11}
because the electronic contributions to the $B_{1g}$ Raman scattering intensity 
are suppressed in the pseudogap regime.\cite{Lin10} 

In addition to high-$T_c$ cuprates, 
electronic nematicity was suggested for the double-layer ruthenates under an applied magnetic field\cite{Borzi07,Rost2009} 
and Fe-based high-$T_c$ superconductors above magnetic phase transitions.\cite{Chu10,Blomberg12} 
A number of theoretical scenarios have been proposed, including 
microscopic phase separation\cite{Honerkamp05,Binz04} and quasi-one-dimensional orbital ordering\cite{Raghu09} for ruthenates, 
and fluctuating magnetic stripe order,\cite{Fang08,Fernandes12} a ferro-orbital ordering,\cite{Kruger09} 
and magnetoelastic coupling\cite{Cano10,Liang11} for Fe superconductors. 
It is worth investigating these scenarios including EL coupling 
with electron correlations treated beyond static mean-field approximations.

Summarizing, 
we investigated the $C_4$ symmetry breaking by the coupling between electrons and the lattice distortion 
using an interacting model for cuprates within CDMFT. 
We found the strong instability towards $C_4$ symmetry breaking in the underdoped pseudogap regime 
in the presence of the strong interaction yielding the Mott transition. 
Thus, this instability is a strong-coupling effect characteristic of a doped Mott insulator. 
Additionally, a weak-coupling instability exists near the van Hove filling, 
but this instability is suppressed by strong correlations 
because the imaginary part of the electron self-energy smears the van Hove singularity. 
On the other hand, the imaginary part of the self-energy plays an essential role for the strong-coupling symmetry breaking 
by increasing the kinetic energy by recovering the quasiparticle coherence. 
This leads to the characteristic Fermi arc reconnection below the pseudogap temperature. 
Our finding can be tested by high-spatial-resolution ARPES.

The authors acknowledge discussions with V. R. Cooper, H.-Y. Kee, and H. Yamase, and 
the support from the U.S. Department of Energy, Basic Energy Sciences, 
Materials Sciences and Engineering Division (S.O.), and 
the Grants-in-Aid for Scientific research, MEXT, Japan (N. F.).


\begin{thebibliography}{*}
\bibitem{footnote}
According to the linear combination of atomic orbitals method (Ref.~\onlinecite{Harrison80}),
the hybridization strength between a $d$ orbital and a $p$ orbital separated by $l$ 
is proportional to $l^{-7/2}$. 
Thus, taking $t \sim 0.5$~eV and the NN Cu bond length $2l \sim 4$~{\AA}, 
a rough estimation for $\alpha$ is $\alpha \approx 7 t/4 \sim 0.9$~eV/{\AA} 
for the ideal orthorhombic distortion $u$.
\bibitem{Note}$F_{ele}$ vs $\delta_t$ analysis becomes unreliable at $N \alt 0.73$ 
because the distance function which emphasizes the low-frequency regime proportional to $1/|\omega_n|$
is not optimal for the overdoped Fermi-liquid regime.\cite{Senechal10} 
The current choice of the distance function results in the increase of $F_{ele}$ with $\delta_t$. 
However, 
it is expected that the phase boundary for finite $U$ will asymptote to that for $U=0$ with decreasing $N$ 
when an appropriate distance function for the Fermi-liquid regime (Ref.~\onlinecite{Senechal10}) or 
a more reliable impurity solver such as the continuous-time quantum Monte Carlo (Ref.~\onlinecite{Gull11}) is employed.
\end{thebibliography}
\end{document}